\documentstyle[preprint,aps,epsf]{revtex}
\tightenlines
\begin{document}

\draft

\title{Spatial distribution of persistent sites}

\author{G.Manoj and P.Ray}

\address{ The Institute of Mathematical Sciences, C. I. T. Campus,
Taramani, Madras 600 113, India}

\date{\today}

\maketitle
\begin{abstract}

We study the distribution of persistent sites (sites unvisited by
particles $A$) in one dimensional   
$A+A\to\emptyset$ reaction-diffusion model. We define the    
{\it empty intervals} as the separations between 
adjacent persistent
sites, and study their size distribution $n(k,t)$ as a function
of interval length $k$ and time $t$.
The decay  of persistence is the process of irreversible coalescence of 
these empty intervals, which we
study analytically under the Independent Interval 
Approximation (IIA). Physical 
considerations suggest that the asymptotic solution is given by the
dynamic scaling form $n(k,t)=s^{-2}f(k/s)$ with the average
interval size $s\sim t^{1/2}$. 
We show under the IIA that the scaling function $f(x)\sim x^{-\tau}$ 
as $x\to 0$ and decays exponentially at large $x$.
The exponent $\tau$ is related to the persistence exponent $\theta$
through the scaling relation $\tau=2(1-\theta)$.
We compare these predictions with the results of numerical simulations.
We determine the two-point correlation function $C(r,t)$ under the
IIA. We find that
for $r\ll s$, $C(r,t)\sim r^{-\alpha}$ where $\alpha=2-\tau$,
 in agreement with our earlier 
numerical results.

\end{abstract}
\pacs{05.40+j, 82.20. Db}

\vspace{1cm}

\section{Introduction}

The persistence of fluctuations in stochastic processes has been an important
topic of study in recent times \cite{SATYA}. Of primary interest in this
context is the persistence probability $P(t)$, which is the
probability that a given stochastic variable $\phi(t)$ with zero mean
retains its sign during the time interval $[0,t]$. A power-law
decay $P(t)\sim t^{-\theta}$ is found in many systems of
physical interest. Consequently, much effort has gone into the
calculation of the new exponent $\theta$ and studying its properties.
There have also been several experimental studies of the persistence
exponent $\theta$ in real life systems\cite{EXPER}.  

A particulary important class of systems whose persistence behaviour has
been investigated are spatially extended systems with a 
stochastic field $\phi({\bf x},t)$ at each lattice site ${\bf x}$.
The time evolution of $\phi({\bf x},t)$ is
coupled to that of its neighbouring sites.
$\phi({\bf x},t)$ could be, for instance, an Ising
spin\cite{BRAY,DERRIDA}, a phase ordering
field\cite{PHORD}, a diffusing field\cite{DIFFUSION} 
or the height of a fluctuating interface\cite{FINT}. 
The persistence probability $P(t)$ is then the fraction of sites ${\bf x}$ where
$\phi({\bf x},t)$ has not flipped sign till time $t$. Recently it was
observed that the set of persistent sites form a fractal and the time 
evolution of their spatial correlations obey dynamic scaling \cite{MANOJ}.
The purpose of this paper
is the investigation of these spatial correlations.

For concreteness, we study the one-dimensional $A+A\to\emptyset$ model. 
Our primary motivation for this choice is
the simplicity of the dynamics of the model which makes an analytic
approach possible. In addition, this model is closely related to the
$d=1$ Glauber-Ising model, which is perhaps the only non-trivial model
where $\theta$ is known exactly. We study  
the distribution of the separations between
nearest neigbour pairs of persistent sites. We call this the Empty
Interval Distribution $n(k,t)$, defined as the number of 
occurences where consecutive persistent sites are  
separated by distance $k$ at time $t$. This quantity is a
direct probe of spatial correlations in the distribution of particles.
If the particles are distributed at random with some average density
$p$, then $n(k)=p^{2}(1-p)^{k}\sim e^{-\lambda k}$ at all $k$.
Any slower mode of decay is indicative of spatial correlations.

In this paper, we study the time evolution of the size distribution
$n(k,t)$ of these Empty Intervals. Persistence decay is 
identified with the
irreversible coalescence of these intervals.
The paper is organised as follows. In the next section we write a 
rate equation for the coalesence of these intervals
under the approximation that the lengths of adjacent intervals are
uncorrelated (IIA). We give phenomenological arguments 
about the asymptotically
relevant dynamical length scale as well as the coalescence probability.
These arguments, combined with the rate equation gives the dynamic
scaling behaviour of $n(k,t)$ at late times $t$. 
We compare our predictions with numerical results. In section III, we
use the IIA to predict the two-point correlations in the distribution of
persistent sites. The predictions are found to be in agreement with 
recent numerical results, showing that IIA is valid.

\section{The Empty Interval Distribution}

In the $A+A\to\emptyset$ model, a set of particles are distributed at
random on the lattice with average density $n_{0}$. Over one time
step, all the particles make an attempted jump to either of the
neighbouring sites with some probability $D$. If two particles meet
each other, both disappear from the lattice. In one dimension, the
density of particles decay with time as $n(t)\sim (8\pi
Dt)^{-\frac{1}{2}}$ as $t\to\infty$\cite{DIFFAN}. 
Persistent sites in $A+A\to\emptyset$
model at any time $t$ are defined as the sites which remained
unvisited by any diffusing particle throughout the time interval $[0:t]$. 
Empty Intervals (which we will call `Interval' for simplicity
henceforth) 
are defined as the separations between two consecutive
persistent sites. By definition, an Interval cannot contain a persistent
site, although it may contain one or more diffusing particles $A$.
The total number (per site) of Intervals of length $k$ at time $t$ is denoted by
$n(k,t)$ and is called the Empty Interval Distribution.

To start with, the the particles are put randomly on the lattice so that
$n(k,t=0)=n_{0}^{2}(1-n_{0})^{k}\sim e^{-\lambda k}$ where $\lambda=-$log$(1-n_{0})$.
With time, the particles diffuse on the lattice, making the sites
non-persistent. $n(k,t)$ evolves satisfying the following
normalisation conditions.
If $I_{m}(t)=
\sum_{k}k^{m}n(k,t)\approx \int_{1}^{\infty}n(s,t)s^{m}ds$ 
is the $m$-th moment of the distribution, then

\begin{equation}
I_{0}(t)=P(t)\sim t^{-\theta}\hspace{0.3cm}; \hspace{0.5cm}
I_{1}(t)=1\hspace{0.5cm} I_{2}(t)\equiv s(t)
\label{eq:NORMAL}
\end{equation}

The first condition follows from the definition of $n(k,t)$, the
second one implies length conservation and the third condition gives 
the mean interval size $s(t)$.
The probability distribution of
interval lengths is $p(k,t)=\frac{n(k,t)}{\sum_{k}n(k,t)}=
P(t)^{-1}n(k,t)$ so that $\sum_{k}p(k,t)=1$.

Two neighbouring Intervals can coalesce when the persistent site between
them is destroyed by a diffusing particle at the boundary of either of
the Intervals. Note that this coalescence process is irreversible.
For simplicity, we consider only binary coalescence in a single time
step where two adjacent Intervals of lengths
$k_{1}$ and $k_{2}$, separated
by a persistent site, coalesce and form a new Interval of length
$k_{1}+k_{2}$ when the persistent site is `killed' by
a particle (Fig. I).
To study this process analytically, we invoke a mean-field
approximation -- the lengths of adjacent Intervals are treated as 
uncorrelated random variables with probability distribution $p(k,t)$. 
This is the Independent Interval Approximation (IIA), which has been used to
study a variety of problems in one dimension\cite{IIA,DIFFAN1}. 

\subsection{Rate Equation for Interval coalescence}

Assuming that IIA is valid, the time evolution of 
$n(k,t)$ is given by the 
rate equation

\begin{eqnarray}
\frac{\partial n(k,t)}{\partial t}=\frac{1}{2}\sum_{m=1}^{k-1}
n(m,t)p(k-m,t)K(m,k-m,t)-\nonumber\\
n(k,t)\sum_{m=1}^{\infty}p(m,t)K(m,k,t)
\label{eq:MASTER}
\end{eqnarray}

where $K(m_{1},m_{2},t)$ is the probability that two adjacent Intervals 
of lengths $m_{1}$ and $m_{2}$ coalesce at time $t$. The first
term in Eq.\ref{eq:MASTER} represents the increase in number of Intervals of size $k$
through coalescence of smaller intervals, while the second term is
the loss term representing the decrease in number when Intervals of
size $k$ merge with other Intervals. 

To solve the above equation for $n(k,t)$, one need to know the form
of the reaction kernel $K(m_{1},m_{2},t)$.
The process
of coalescence of Intervals involves the destruction of the persistent site in
between them by a particle, which can come from either of the Intervals.
So, quite generally,

\begin{equation} 
K(m_{1},m_{2},t)=Q(m_{1},t)+Q(m_{2},t)
\label{eq:SPLIT}
\end{equation}

where $Q(m,t)$ is the fraction of intervals of size $m$ which is
destroyed at time $t$. $Q(m,t)$ satisfies the following condition by definition.

\begin{equation}
\sum_{m}n(m,t)Q(m,t)=-\frac{\partial P(t)}{\partial t}=\frac{\theta}{t}P(t)
\label{eq:ALPHA}
\end{equation}

where we have made use of the fact that $P(t)\sim t^{-\theta}$.

The form of $Q(m,t)$ can be argued for in the following way.
An Interval
of length $m$ at time $t$ can contain a particle anywhere inside it 
only if the interval length is at least of the order of the diffusive
scale $\sqrt{Dt}$. That is, $Q(m,t)\simeq 0$ for $m\ll \sqrt{Dt}$.  
It is also known that the particle distribution is correlated
over length scales $r\ll \sqrt{Dt}$, wheras it is completely random
over $r\gg \sqrt{Dt}$\cite{DIFFAN1}. So we expect that for $m\gg \sqrt{Dt}$, 
$Q(m,t)\to \alpha(t)$, independent of $m$.
These physical considerations leads us to suggest the following
dynamic scaling form for $Q(m,t)$.

\begin{equation}
Q(m,t)=\alpha(t)\beta(\frac{m}{\sqrt{Dt}})
\label{eq:SCALQ}
\end{equation}

where the function $\beta(x)$ is expected to have a sigmoidal form,
ie., $\beta(x)=0$ for $x\ll 1$ and $\beta(x)\to 1$ for $x\gg 1$.
The function $\alpha(t)$ will be determined later.

\subsection{Dynamic scaling}

We assume that at asymptotic times, the
distribution $n(k,t)$ is characterised by a single dynamic 
length scale $s(t)$. We note that there are
two relevant length scales in the problem.
The first is the diffusive scale ${\cal L}_{D}(t)\sim \sqrt{Dt}$ entering the
scaling form Eq.\ref{eq:SCALQ} for the coalescence probability. 
On the other hand, the inverse of the persistent fraction $P(t)$ is
also a length scale, which we shall call the persistence scale,
denoted by ${\cal L}_{p}(t)\sim t^{\theta}$. The asymptotic behaviour
is expected to be dominated by the larger of the two, ie., the
diffusive scale ${\cal L}_{D}(t)$ in the present case (since $\theta < 1/2$).

We now invoke the dynamic scaling
ansatz, ie., $n(k,t)\propto f(\frac{k}{s})$
with 
\begin{equation}
s\sim t^{1/z},\: z=2
\label{eq:DYNSCAL}
\end{equation}

From the length conservation
condition given by the second part of Eq. \ref{eq:NORMAL} it follows
that the prefactor is $\sim s^{-2}$.
Thus, the scaling solution for $n(k,t)$ is written in the form

\begin{equation}
n(k,t)=s(t)^{-2}f\left(\frac{k}{s(t)}\right)
\label{eq:SCAL}
\end{equation}

Substituting Eq. \ref{eq:SCAL} and Eq. \ref{eq:SCALQ} in
Eq. \ref{eq:ALPHA}, we find $\alpha(t)$.

\begin{equation}
\alpha(t)=\frac{\theta}{t}\frac{s(t) P(t)}{B}
\label{eq:ALPHA1}
\end{equation}

where $B=\int_{0}^{\infty}\beta(x)f(x)dx$. Substituting
Eq.\ref{eq:SCAL} in the normalisation conditions Eq.\ref{eq:NORMAL}, we find the
following conditions on the scaling function.

\begin{equation}
\int_{s^{-1}}^{\infty}f(x)dx=s P(t)\hspace{0.3cm}; \hspace{0.5cm} \int_{0}^{\infty}f(x)xdx=1
\label{eq:NORMAL1}
\end{equation}

In the first integral, the lower limit is set as $s(t)^{-1}$ 
to take care of any possible small argument divergence.

Substituting Eq.\ref{eq:SCALQ}, \ref{eq:SCAL}, \ref{eq:ALPHA1} and \ref{eq:DYNSCAL} 
in Eq. \ref{eq:MASTER}, we find the following equation for
the scaling function $f(x)$.

\begin{eqnarray}
\frac{\eta}{z}\frac{\partial f}{\partial\eta}=-\frac{\theta}{B}\int_{s(t)^{-1}}^{\frac{\eta}{2}}
f(x)f(\eta-x)[\beta(x)+\beta(\eta-x)]dx -\nonumber\\
\left[\frac{2}{z}-\theta-\frac{\theta}{B}s(t) P(t)\beta(\eta)\right]f(\eta)
\label{eq:MASTERSCAL}
\end{eqnarray}

where the scaling variable $\eta=\frac{k}{s(t)}$. 

{\it Case} I: $\eta\ll 1$.

For $\eta\ll 1$, all $\beta(x)\simeq 0$ for $x\leq \eta$. 
This case corresponds to small Intervals, ie., those which are not
large enough to contain a diffusing particle till time $t$.
In this case, the equation reduces to 
$\eta\frac{\partial f}{\partial\eta}=-(2-z\theta)f(\eta)$ which has
the solution $f(\eta)\sim \eta^{-\tau}$ where the new exponent $\tau$
is related to $\theta$ through the scaling relation 

\begin{equation}
\tau=2-z\theta
\label{eq:SCALREL}
\end{equation}

From Eq. \ref{eq:SCAL} this implies that for $k\ll s$, 
$n(k,t)\sim t^{-\theta}k^{-\tau}$. 
For the model under consideration here, $\theta$ is known exactly 
to be $3/8$ \cite{DERRIDA} which gives $\tau = 5/4$.

{\it Case} II: $\eta\gg 1$.

For general values of $\eta$, $\beta(\eta)$ is non-zero, and because
$\tau> 1$, the first integral diverges near $x=0$ as $x^{-(\tau-1)}$.
There is another divergence in the last term, of the form
$t^{1/z-\theta}$. It can be shown that this term can be exactly cancelled by
the divergent part of the first integral.
After carrying out this `regularisation' (details to be found in 
Appendix A) and putting $z=2$ in Eq.\ref{eq:MASTERSCAL} the equation for the 
scaling function $f(\eta)$ stands as

\begin{eqnarray}
\frac{\eta}{2}\frac{\partial f}{\partial\eta}=
-\frac{\theta}{B}\int_{0}^{\frac{\eta}{2}}
f(x)f(\eta-x)[\beta(x)+\beta(\eta-x)-\beta(\eta)]dx-\nonumber \\
\frac{\theta}{B}
\beta(\eta)\int_{0}^{\frac{\eta}{2}}f(x)[f(\eta-x)-f(\eta)]dx-
\left[1-\theta-\frac{\theta}{B}\beta(\eta)\int_{\frac{\eta}{2}}^{\infty}f(x)dx\right]
f(\eta)
\label{eq:MASTER-REG}
\end{eqnarray}

A general solution of this equation requires the knowledge of
the detailed form of the scaling function $\beta(\eta)$. However, for
large values of $\eta$ where $\beta(\eta)\simeq 1$, one can simplify this
equation. We define the point $\eta^{*}$ sufficiently large such
that for $\eta\geq \eta^{*}$, $\beta(x)=1$ within the limits of
accuracy required. Without any loss of generality, one can put
$\eta^{*}=1$ by rescaling the length scale $s(t)$ accordingly. 
For $\eta\geq 1$, we define $f(\eta)\equiv h(\eta)$, whose equation is

\begin{eqnarray}
\frac{\eta}{2}\frac{\partial h}{\partial \eta}=
-\frac{\theta}{B}\left[2\int_{1}^{\frac{\eta}{2}}h(x)h(\eta-x)dx+\int_{0}^{1}f(x)
[h(\eta-x)-h(\eta)]dx\right]-(1-2\theta)h(\eta)
\label{eq:MASTER-EXP}
\end{eqnarray}

This equation has 
a solution of the form $h(\eta)=Ge^{-\lambda\eta}$ as can be shown by direct substitution.
The constants $G$ and $\lambda$ are related through the relations

\begin{equation}
\lambda B=2\theta G
\label{eq:ALPS1}
\end{equation}

and

\begin{equation}
\lambda+2\theta=1+\frac{\theta}{B}F(\lambda)
\label{eq:ALPS2}
\end{equation}

where $F(\lambda)=\int_{0}^{1}f(x)\left[e^{\lambda x}(1+\beta(x))-1\right]dx$
and 
\begin{equation}
B=\int_{0}^{1}f(x)\beta(x)dx+\frac{G}{\lambda}e^{-\lambda}
\label{eq:ALPS3} 
\end{equation}

by definition. Eq.\ref{eq:ALPS1}-\ref{eq:ALPS3} formally gives the
constants $\lambda$ and $G$. However, the actual evaluation of these
constants requires the knowledge of the function $f(x)$ in the entire
range [0:1] (and not just near $x=0$, where $f(x)\sim x^{-\tau}$),
which, in turn, is possible only if the detailed form of $\beta(x)$ is
known. Hence we will restrict ourselves to showing that the parameter
$\lambda > 0$, which is required for the solution to be physically reasonable.

In Eq. \ref{eq:ALPS3}, we note that $B\geq
\frac{G}{\lambda}e^{-\lambda}$, depending on how sharply 
$\beta(x)$ rises near $x=1$. The equality holds for the step
function $\beta(x)=\Theta(x-1)$ where $\Theta(x)=0$ for $x<0$ and
$\Theta(x)=1$ for $x\geq 1$.
After using this inequality in Eq.\ref{eq:ALPS1}, we find that
$\lambda\geq -$log$(2\theta)$. Since $\theta< 1/2$, it follows that
$\lambda > 0$.

\subsection{Numerical Results}

We determine the distribution $n(k,t)$ 
numerically by simulating $A+A\to\emptyset$ model on 
one dimensional lattice 
of size $N= 10^{5}$ with periodic boundary condition.
Particles are initially distributed at random on the lattice with some
average density $n_{0}$, and
their positions are sequentially updated--- each particle
is made to move one step in either direction with probability 
$D=1/2$. When two particles meet each other, both are removed
from the lattice. The time evolution is observed 
up to $10^{5}$ Monte-Carlo steps (1 MC step is counted after all the
particles in the lattice were touched once). The simulation is repeated
for several random starting configurations of the particles    
for any particular initial density and we repeat the entire 
simulation for four different initial density $n_0$. For any $n_0$,   
we determine the number of intervals of length $k$ (per site) at
time $t$. 

To compute the mean interval size $s(t)$, we ran the
simulation upto $t=10^{5}$ time steps, and averaged the results over
100 starting distributions of particles, with the same initial density.
In Fig. II, we plot $s(t)$ vs $t$ for four different values of $n_{0}$--
0.2, 0.5, 0.8 and 0.95. 
For $n_{0}=1/2$, we find that $s(t) \sim at^{1/z}$ with $z\simeq 1.97(1)$
and $a\simeq 5.96$. 
But for other values of $n_0$, we find that the observed value of $z$ is
different from 2. 
In Fig. III, the running exponent $d($log$ s)/d($log$ t)$ is plotted 
with $1/($log$ t)$ and the results show the systematic deviation away from the 
value $1/2$ expected 
from the scaling picture presented in the previous section. We will 
discuss about the possible origin of this deviation later.

In Fig. IV, we plot the scaling function 
$f(x)=s(t)^{2}n(k,t)$ against the scaling variable
$x=k/s(t)$ for $t=10^{4}$ and $3.10^{4}$. To get the nature of the scaling 
function one needs to average over lots of configurations. This has restricted 
us to smaller time steps and data for three values of
initial density, $n_{0}=0.2, 0.5$ and $0.8$ averaged over 500, 1000 
and 1500 different initial distribution of particles respectively. 
For all $n_0$, we
find that the scaling function $f(x) \sim x^{-\tau}$ for
$x << 1$ and decays exponentially for higher values of $x$.
For $n_{0}=0.5$, we find $\tau=1.25(1)$ in accordance with the scaling 
relation Eq \ref{eq:SCALREL}. For $n_{0}=0.2$, we
find $\tau\simeq 1.32(2)$ while for $n_{0}=0.8$, the observed value of
$\tau$ is 1.13(2). For all $n_0$, $\tau$ satisfies the scaling relation 
Eq. \ref{eq:SCALREL} if $z$ is replaced by its effective value.

For general values of $n_{0}$, we find that the numerical values of $s(t)$ 
supports the following form (within the time range studied)  

\begin{equation} 
s_{n_{0}}(t)\sim at^{1/z}+b(n_{0})t^{\phi}.  
\label{eq:DELTSCAL}
\end{equation}

The non-universal constant $b$ is $<, =$ or $>0$ for $n_{0}<$, $=$ or $>0.5$.  
To compute the prefactor $b$ and the exponent $\phi$, 
we plot the difference $\Delta s_{n_{0}}(t)=|s_{n_{0}}(t)-s_{1/2}(t)|$ vs $t$,
for $n_{0}=0.2,0.8$ and 0.95 (Fig. V). The exponent $\phi$ 
is numerically found to be close to the persistence
exponent $\theta = 0.375$ (Table I). 
We find that as $n_{0}\to 1$, the constant $b$
undergoes a sharp rise so that the effective dynamical exponent of $s(t)$
is numerically close to $\theta$ for an appreciable range in time
(Fig. III).
At the same time, we note that only the first term in Eq.\ref{eq:DELTSCAL}
is asymptotically relevant since $\phi< \frac{1}{2}$. 

The two terms in Eq.\ref{eq:DELTSCAL} can have their origin from the 
two dynamical length scales in the problem, the diffusive scale 
${\cal L}_D(t) \sim t^{1/2}$ and the persistence scale ${\cal L}_p(t) \sim t^{\theta}$. 
For large $n_0$, the typical interval length between two consecutive 
persistent sites is determined by the decay of persistence only, rather than 
the diffusion of the particles. So, it is understandable that the 
dynamical behavior of $s(t)$ coincides with that of ${\cal L}_p(t)$ at least 
at the initial times. However at late times, when the particle density 
falls down as a result of annihilation, the situation becomes same as that 
of starting with low $n_0$ and the decisive scale is ${\cal L}_D(t)$. However, 
the precise form and behavior of the prefactor $b(n_0)$ with $n_0$ 
remains to be understood.

\section{Two-point correlations}

A good picture of the spatial distribution of the persistent sites and the 
presence of any possible correlation in their distribution is obtained 
from the two-point correlation   
$C(r,t)$, which is defined as
the probability that site {\bf x+r} is persistent, given that site
{\bf x} is persistent (averaged over {\bf x}).

\begin{equation}
C(r,t)= \langle\rho({\bf x},t)\rangle^{-1}\langle\rho({\bf x},t)\rho({\bf x+r},t)\rangle
\label{eq:CORR}
\end{equation}

where the brackets denote average over the entire lattice and
$\rho({\bf x},t)$ is the density of persistent sites: 
ie., $\rho({\bf x},t)=1$ if site ${\bf x}$ is persistent at time $t$, and 0 otherwise. 
Clearly, $\langle\rho({\bf x},t)\rangle=P(t)$ by definition.

Within the IIA, the relation between $C(r,t)$ and $n(r,t)$ (We
consider $r\gg
1$, so that the discreteness of the underlying lattice can be ignored)
can be written as the following infinite series:

\begin{eqnarray}
C(r,t)=P(t)^{-1}n(r,t)+
P(t)^{-2}\int_{1}^{r}dx\: n(x,t)n(r-x,t)+
\nonumber \\
P(t)^{-3}\int_{1}^{r}dx \:n(x,t)\int_{1}^{r-x}dy \:n(y,t)n(r-x-y,t)+....
\label{eq:REAL1}
\end{eqnarray}

The first term corresponds to the case where there is no other
persistent site in the range $[0:r]$, ie., a single Interval
of length $r$. The second term gives the probability that
the range is split into two Intervals of length $x$ and $r-x$ by the
presence
of a persistent site at $x$, the third term gives the probability for three Intervals and so on.

The above series can be rewritten as the following self-consistent
equation for $C(r,t)$.

\begin{equation}
P(t)C(r,t)=n(r,t)+\int_{1}^{r}n(x,t)C(r-x,t)dx
\label{eq:REAL}
\end{equation}

In terms of the Laplace transforms ${\tilde C}(p,t)=\int_{1}^{\infty}C(r,t)e^{-pr}dr$
and ${\tilde n}(p,t)=\int_{1}^{\infty}n(s,t)e^{-ps}ds$ Eq. \ref{eq:REAL}
becomes  

\begin{equation}
{\tilde C}(p,t)=\frac{{\tilde n}(p,t)}{P(t)-{\tilde n}(p,t)}
\label{eq:LAPLACE}
\end{equation}

From Eq.\ref{eq:SCAL}, we find 
\begin{equation}
{\tilde n}(p,t)=s^{-1}{\tilde f}(ps) 
\label{eq:FTILD}
\end{equation}

where ${\tilde f}(q)=\int_{s^{-1}}^{\infty}f(\eta)e^{-q\eta}d\eta$.
which can be written in the following regularised form, using Eq.\ref{eq:NORMAL1}.

\begin{equation}
{\tilde f}(q)=s(t)P(t)-f_{1}(q) 
\label{eq:NEW}
\end{equation}

where
\begin{equation}
f_{1}(q)=\int_{0}^{\infty}f(\eta)[1-e^{-q\eta}]d\eta. 
\label{eq:F1Q}
\end{equation}

Substituting Eq.\ref{eq:FTILD}, \ref{eq:NEW} and Eq.\ref{eq:F1Q} into
Eq.\ref{eq:LAPLACE} we find that 

\begin{equation}
{\tilde C}(p,t)=\frac{s(t)P(t)}{f_{1}(ps)}-1
\label{eq:NEW1}
\end{equation} 

The second term in RHS can be neglected at late times, since
$s(t)P(t)$ diverges as $t^{1/z-\theta}$. It follows that 
in this limit, $C(r,t)$ has the dynamic scaling form


\begin{equation}
C(r,t)=P(t)g\left(\frac{r}{s(t)}\right)
\label{eq:CORSCAL}
\end{equation}

where 

\begin{equation}
{\tilde g}(q)=\frac{1}{f_{1}(q)}
\label{eq:IMP}
\end{equation}

is the Laplace transform of $g(x)$: ${\tilde
g}(q)=\int_{0}^{\infty}g(x)e^{-qx}dx$. 

The preceding expressions can be used to deduce the limiting behaviour
of the scaling function $g(\eta)$ for the cases $\eta\ll 1$
and $\eta \gg 1$, without needing to solve Eq.\ref{eq:REAL1} or \ref{eq:REAL}
explicitly.

{\it Case} I: $\eta\gg 1$.

To find the asymptotic behaviour of $g(\eta)$, we note that $f_{1}(q)$
vanishes near $q=0$ as $f_{1}(q)\sim q$. Thus ${\tilde g}(q)\sim \frac{1}{q}$ as
$q\to 0$ from Eq.\ref{eq:IMP}. By standard results in the theory of 
Laplace transforms \cite{LAPLACE},
this implies that $g(\eta)\sim 1$ as $\eta\to \infty$.

{\it Case} II: $\eta\ll 1$.

To analyse this case, consider the real-space relation Eq.\ref{eq:REAL1}.
For $\eta\ll 1$, or equivalently, $r\ll s$, we have 
shown that $n(r,t)\sim P(t)r^{-\tau}$.
It is clear that in this range, the RHS of Eq.\ref{eq:REAL1} 
is time independent, so $C(r,t)$ in the LHS should also be time independent.
From the dynamic scaling form Eq.\ref{eq:CORSCAL}, we find that this
is possible only if the scaling function is a power-law near
the origin: $g(\eta)\sim \eta^{-\alpha}$ as $\eta\to 0$. After
substituting in Eq.\ref{eq:CORSCAL} and requiring the resulting
expression to be time independent, we find 

\begin{equation}
\alpha=z\theta
\label{eq:SCALREL1}
\end{equation}

We find $C(r,t)\sim r^{-\alpha}$ for $r\ll s$ and $C(r,t)\simeq
P(t)$ for $r\gg s$. The power law decay at small distances is 
expected, because the RHS of Eq.\ref{eq:REAL1} contains only scale
invariant terms in this limit, hence the LHS also should be likewise.
In Appendix B, we show that this is also consistent with Eq.\ref{eq:REAL}.

We see that in the IIA calculation, the length scale $s(t)$
demarcates the correlated and uncorrelated regions for $C(r,t)$.
In the correlated region ($r \ll s(t)$), the persistent sites form 
a fractal with fractal dimension $d_f = d - \alpha = \frac{1}{4}$, with the 
correlation length $s(t)$ increasing with time as $s \sim t^{1/2}$.  
The IIA results agree very well with that of numerical simulations
\cite{MANOJ}, showing the validity of the approximation.  

\section{Conclusion}

Persistent sites are shown to have strong correlations in their 
spatial distribution. In one dimensional $A+A \to\emptyset$ reaction-diffusion 
system, we show that there is a length scale $s(t)$, diverging with 
time as $s(t) \sim t^{1/z}$, which demarcates the correlated 
region from the uncorrelated one. We argue that $z=2$ at large $t$
limit. Persistent sites 
separated by distance $k\ll s(t)$ are highly unlikely to have a 
particle $A$ between them and so retains their persistent character. 
Only persistent sites separated by distance $\gg s(t)$ 
take part in the decay of persistence at subsequent times. 

We find that if $k$ is the distance of separation between any two 
consecutive persistent sites, then for   
$k\ll s(t)$, the distribution of $k$ is scale-free and  decays 
algebraically as $k^{-\tau}$ with $\tau = 2 -z\theta$. 
We show this using the IIA (Independent 
Interval Approximation), which assumes no correlation in 
the lengths of any two adjacent intervals. 
We have verified our results by 
numerical simulations which suggests the validity of the IIA.   
Under the IIA, our calculation 
for the two-point correlation shows that over length scales $r \ll s(t)$,  
the persistent site distribution over the lattice is a fractal  
with dimension $d_f = \tau-1$, in accordance to our earlier observations 
\cite{MANOJ}.  

\section{ACKNOWLEDGEMENTS}
We thank G. I. Menon for a critical reading of the manuscript and
valuable suggestions.

\begin{appendix}
\section{}
The divergence in the first integral in Eq. \ref{eq:MASTERSCAL} can be separated
out as follows. We write
$f(\eta-x)=f(\eta)+\Delta_{x}f(\eta)$ and
$\beta(\eta-x)=\beta(\eta)+\Delta_{x}\beta(\eta)$.
so that $\lim_{x\to 0}\Delta_{x}f(\eta)=\lim_{x\to 0}\Delta_{x}\beta(\eta)=0$.

After substituting for $f(\eta-x)$ and $\beta(\eta-x)$, the divergent
part of the integral separates into the following terms.

\begin{eqnarray*} 
\int_{s(t)^{-1}}^{\frac{\eta}{2}}f(x)f(\eta-x)\beta(\eta-x)dx=
f(\eta)\beta(\eta)\int_{s(t)^{-1}}^{\frac{\eta}{2}}f(x)dx+
f(\eta)\int_{0}^{\frac{\eta}{2}}f(x)\Delta_{x}\beta(\eta)dx+\\
\beta(\eta)\int_{0}^{\frac{\eta}{2}}f(x)\Delta_{x}f(\eta)dx+\int_{0}^{\frac{\eta}{2}}
f(x)\Delta_{x}f(\eta)\Delta_{x}\beta(\eta)dx
\end{eqnarray*}

The first term is divergent near the origin, while all other terms are
finite by construction. Now we rewrite the first term using the
equality $\int_{s(t)^{-1}}^{\infty}f(x)dx=s(t)P(t)$. After some
simplifications, the integral becomes

\begin{eqnarray*}
\int_{s(t)^{-1}}^{\frac{\eta}{2}}f(x)f(\eta-x)\beta(\eta-x)dx=
f(\eta)\beta(\eta)s(t)P(t)+\int_{0}^{\frac{\eta}{2}}f(x)f(\eta-x)\left[\beta(\eta-x)-
\beta(\eta)\right]dx+\\
\int_{0}^{\frac{\eta}{2}}f(x)\left[f(\eta-x)-f(\eta)\right]-
\beta(\eta)f(\eta)\int_{\frac{\eta}{2}}^{\infty}f(x)dx
\end{eqnarray*}

The first term is the divergent part of the integral, which exactly
cancels the last term in Eq. \ref{eq:MASTERSCAL}, to give the regularised
Eq. \ref{eq:MASTER-REG}.

\section{}

For $r\gg 1$, it is reasonable to assume that the higher
order terms in the RHS of Eq.\ref{eq:REAL1} will contribute more than the first term, ie.,
the range $[0:r]$ is more likely to be covered with more than
one Interval than a single one of length $r$. After using this
approximation, and substituting $n(r,t)\simeq (\tau-1)P(t)r^{-\tau}$
in the continuum limit,
Eq. \ref{eq:REAL} is simplified to
\[
C(r,t)\simeq (\tau-1)\int_{1}^{r-1}(r-x)^{-\tau}C(x,t)dx 
\]

Our purpose is to see if the equation
\begin{equation}
r^{-\alpha}\simeq (\tau-1)\int_{1}^{r-1}x^{-\alpha}(r-x)^{-\tau}dx
\label{eq:B1}
\end{equation}

is consistent for $\alpha=z\theta=2-\tau$ (Eq.\ref{eq:SCALREL}) at $r\gg 1$. 

The integral $I=\int_{1}^{r-1}x^{-\tau}(r-x)^{-\alpha}dx$
can be transformed by change of variables into the more standard
form\cite{INTEGRAL} 
\[\int_{0}^{r-2}(1+y)^{-\tau}[r-1-y]^{-\alpha}dy \simeq \frac{r^{1-\alpha}}{1-\alpha}
F(1, \tau; 2-\alpha; -r)\hspace{0.3cm}$for$\hspace{0.2cm}\alpha< 1 \hspace{0.2cm}$and
$\hspace{0.2cm}r\gg 1. \]

where $F(a, b ; c ; z)$ is the Gauss Hypergeometric function.
For $b=c$, $F(a,b;b;z)=(1-z)^{-a}$ exactly, independent
of $b$\cite{GAUSS}. Thus, for $\alpha=2-\tau$ we find

\[
(\tau-1)I= r^{-\alpha}[1+o(\frac{1}{r})]
\]
which is consistent with Eq. \ref{eq:B1}, at $r\gg 1$.

\end{appendix}

\begin{table}
\begin{tabular}{ccc}
$n_{0}$ & $b$  & $\phi$\\
\hline
0.20 & $-6.621$ & 0.34372(11)\\
0.80 & 15.701 & 0.35495(5) \\
0.95 & 84.672 & 0.36572(4) \\
\end{tabular}
\narrowtext
\vspace{0.5cm}
\caption { 
Results for the prefactor $b$ and exponent $\phi$ as measured from
simulations. The numerical value of $\phi$ is found to be close to
the persistence exponent $\theta$ whose exact value is 0.375. The
figures in brackets represent statistical error in the last decimal
place. Note the sharp rise in $b$ as $n_{0}\to 1$.
}
\label{tab:TAB1}
\end{table}

\begin{figure}[a]
\narrowtext
\epsfxsize=1.5in
\epsfbox{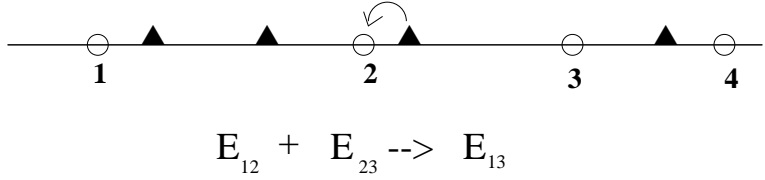}
\vspace{-10.0cm}
\caption
{
In the picture, white circles are persistent sites (numbered 1, 2,
3..) 
and dark triangles are diffusing particles.
Two Empty Intervals $E_{12}$ and $E_{23}$ are shown to merge
together to give a new Interval $E_{13}$ when the persistent site 2 
at the boundary is killed by a diffusing particle.
}
\end{figure}

\begin{figure}[b]
\narrowtext
\epsfxsize=2.5in
\epsfbox{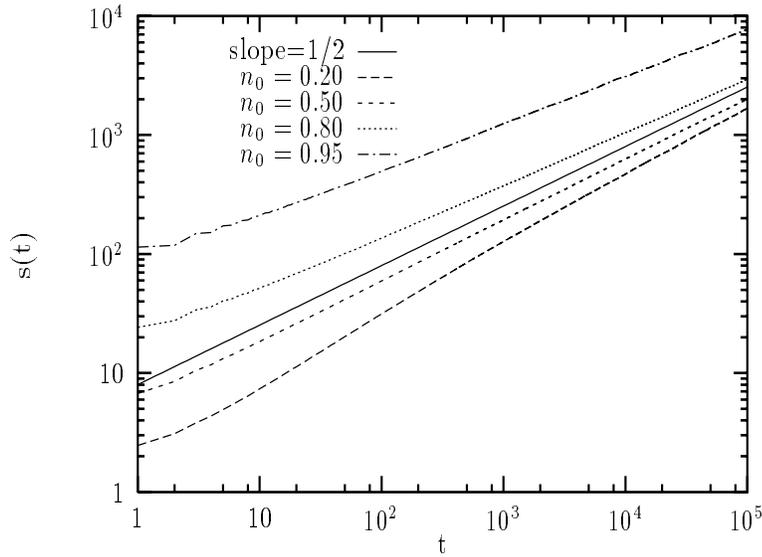}
\caption
{
The length scale $s(t)$ is plotted as a function of time $t$. The
straight line is a fit, with slope 1/2. 
}
\end{figure}

\begin{figure}[c]
\narrowtext
\epsfxsize=2.5in
\epsfbox{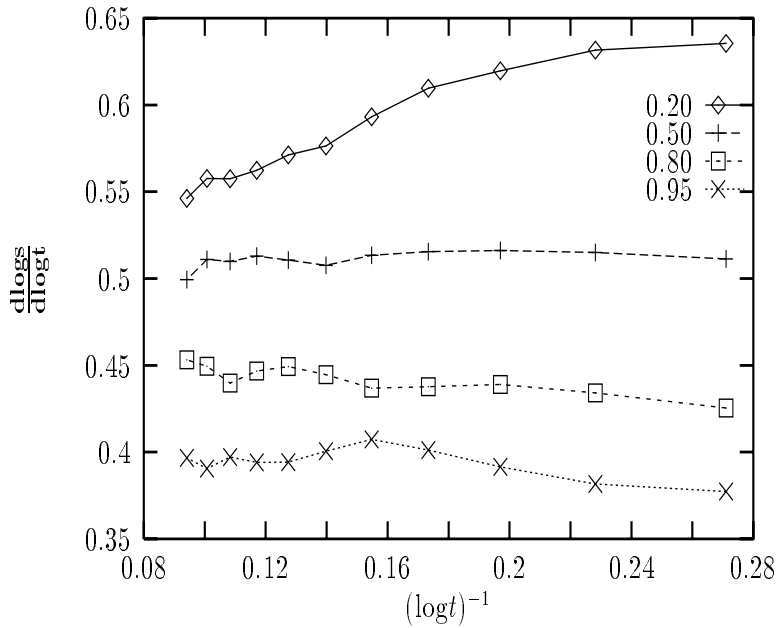}
\vspace{-1.0cm}
\caption
{
The effective exponent $\frac {d logs}{d logt}$ is plotte against
$1/$log$ t$ for four values of starting density. For $n_{0}=0.5$,
the exponent value is close to 0.5, expected from the scaling
arguments. For other values of $n_{0}$, systematic deviations away
from 0.5 is observed.}
\end{figure}

\begin{figure}[d]
\narrowtext
\epsfxsize=2.5in
\epsfbox{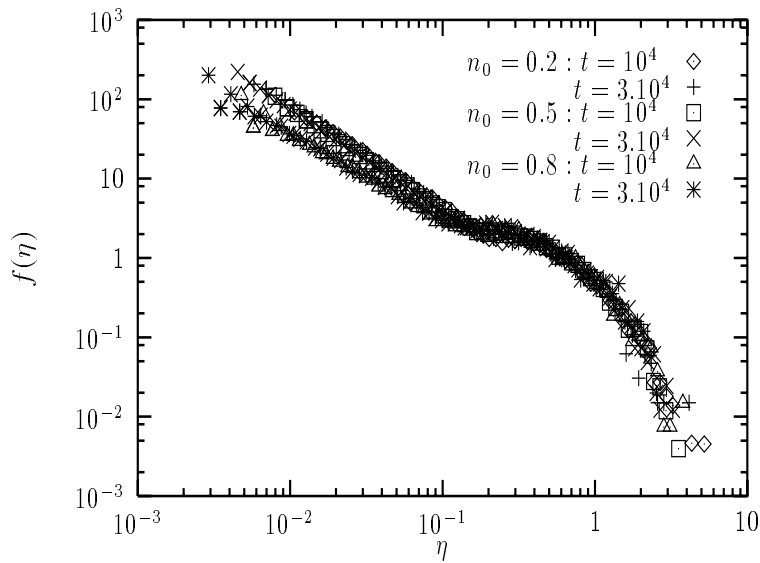}
\caption
{
The scaling function $f(\eta)=s(t)^{2}n(k,t)$ is plotted against
the scaling variable $\eta=k/s(t)$ on a logarithmic scale.
There is a power-law divergence at small $\eta$ and exponential decay
at large $\eta$, as predicted by the IIA calculation. The observed
value of $\tau$ for $n_{0}=0.8$ is seen to be appreciably
different from that for other $n_{0}$.
}
\end{figure}

\begin{figure}[e]
\narrowtext
\epsfxsize=2.5in
\epsfbox{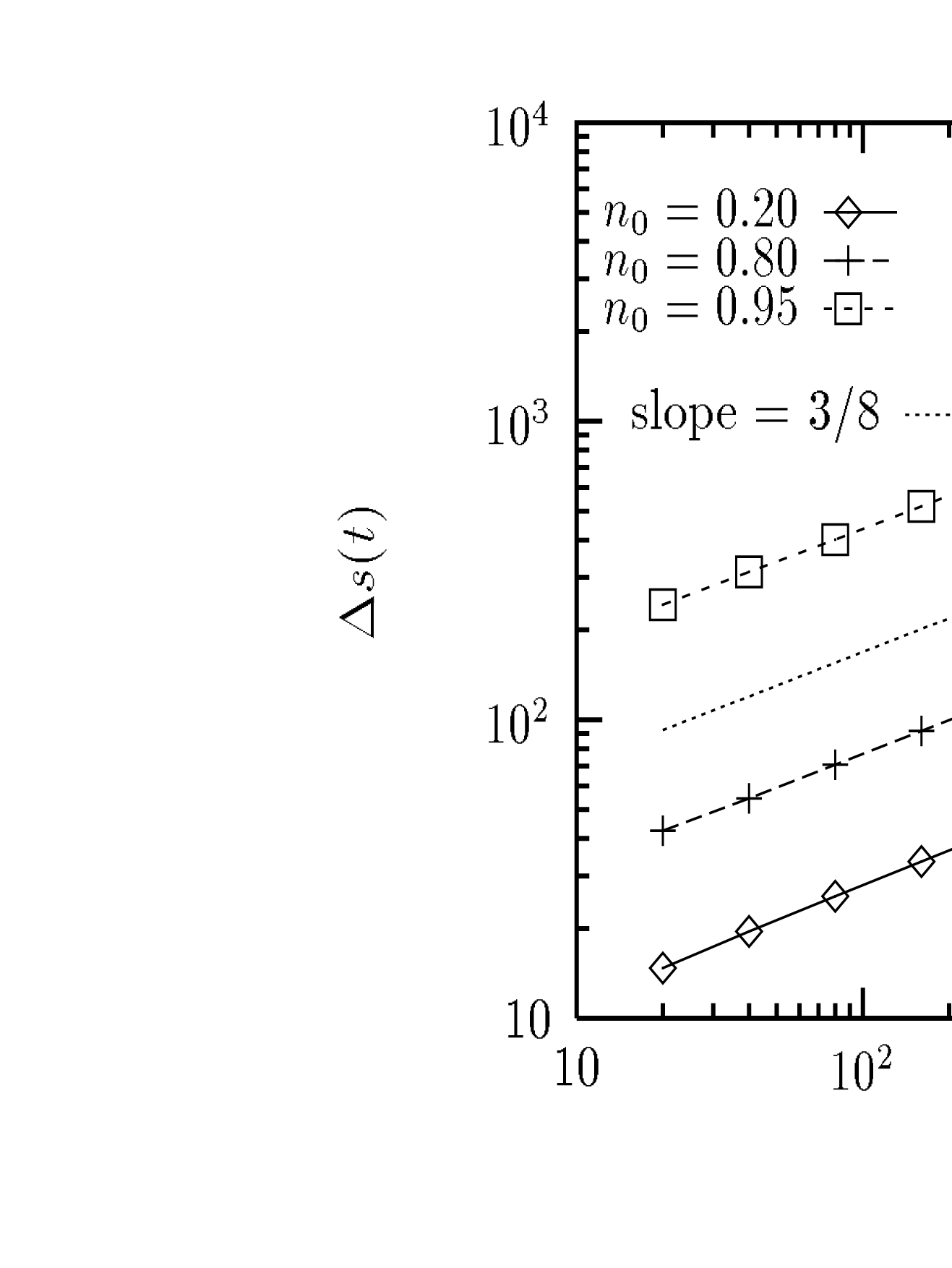}
\caption
{
The difference $\Delta s(t)=|s(t)-s_{1/2}(t)|$ is plotted against $t$
for $n_{0}=0.2,0.8$ and 0.95. The straight line is a fit with
slope 3/8.
}
\end{figure}


\begin{references}

\bibitem{SATYA} For a recent review, S. N. Majumdar, cond-mat/9907407.

\bibitem{EXPER} M. Marcos-Martin {\it et. al}, Physica A {\bf 214},
396 (1995); W. Y. Tam {\it et. al}, Phys. Rev. Lett. {\bf 78}, 1588 (1997).

\bibitem{BRAY} B. Derrida, A. J. Bray and C. Godr\`{e}che, J. Phys. A
{\bf 27}, L357 (1994); D. Stauffer, J. Phys. A {\bf 27}, 5029 (1994).

\bibitem{DERRIDA} B. Derrida, V. Hakim and V. Pasquier, Phys. Rev. Lett.
{\bf 75}, 751 (1995), and J. Stat. Phys. {\bf 85}, 763 (1996). 
	
\bibitem{PHORD} A. J. Bray, B. Derrida and C. Godr\`{e}che, Europhys. Lett.
{\bf 27}, 175 (1994).

\bibitem{DIFFUSION} S. N. Majumdar, C. Sire, A. J. Bray and S. J. Cornell,
Phys. Rev. Lett {\bf 77}, 2867 (1996); B. Derrida, V. Hakim and R. Zeitak,
Phys. Rev. Lett {\bf 77}, 2971 (1996).

\bibitem{FINT} J. Krug, H. Kallabis, S. N. Majumdar, S. J. Cornell,
A. J. Bray and C. Sire, Phys. Rev. E {\bf 56}, 2702 (1997);
H. Kallabis and J. Krug, preprint: cond-mat/9809241

\bibitem{MANOJ} G. Manoj and P. Ray, cond-mat/9912209, to appear 
in J. Phys. A.

\bibitem{DIFFAN} D. Toussaint and F. Wilczek, J. Chem. Phys. {\bf 78},
2642 (1983); 
D. C. Torney and H. M. McConnell, J. Phys. Chem. {\bf 87},
1941 (1983);
A. A. Lushnikov, Phys. Lett. A {\bf 120}, 135 (1987);
J. L. Spouge, Phys. Rev. Lett. {\bf 60}, 871 (1988).

\bibitem{IIA} B. Derrida, C. Godr\`{e}che and Y. Yekutieli, Phys. Rev. A
{\bf 44} 6241 (1991). 

\bibitem{DIFFAN1} P. A. Alemany, D. ben-Avraham, Phys. Lett. A
{\bf 206}, 18 (1995).

\bibitem{LAPLACE} M. G. Smith, {\it Laplace Transform Theory},
D. Van Nostrand Company, London (1966).



\bibitem{INTEGRAL} {\it Tables of Integral Transforms},
ed. A. Erd\`{e}lyi (McGraw Hill, 1954).

\bibitem{GAUSS}  {\it Handbook of Mathematical Functions}, ed. M. Abramovitz and
I. A. Stegun (Dover, New York, 1968).

\end{references}
\end{document}